\documentclass[amsmath,amssymb,aps,prd]{revtex4}

\newcommand {\slsh} [1] {\not{\hbox{\kern-2pt${#1}$}}}

\begin{document}

\preprint{CERN-PH-TH-2012-095}

\title{QCD With A Chemical Potential, Topology,\\
And The 't Hooft $1/N$ Expansion}

\author{Adi Armoni}
\email{a.armoni@swan.ac.uk}
\affiliation{Department of Physics, Swansea University\\ Singleton Park, Swansea, SA2 8PP, UK}

\author{Agostino Patella}%
\email{agostino.patella@cern.ch}
\affiliation{Physics Department, CERN\\1211 Geneva 23, Switzerland}

\date{\today}

\begin{abstract}
We discuss the dependence of observables on the chemical
potential in 't Hooft's large-$N$ QCD.
To this end, we use the worldline formalism to expand the fermionic
determinant in powers of $1/N$. We consider the hadronic as well as
the deconfining phase of the theory. We discuss the origin of the
sign problem in the worldline approach and elaborate on the planar
equivalence between QCD with a baryon chemical potential and QCD with
an isospin chemical potential. We show that for C-even observables the sign problem occurs at a
subleading order in the $1/N$ expansion of the fermionic determinant. Finally, we comment on the finite $N$ theory.

\bigskip

CERN-PH-TH-2012-095
\end{abstract}

\maketitle

\section{Introduction}

Understanding the QCD phase diagram is an important and longstanding problem (see Ref. \cite{Stephanov:2007fk} for a review). Since the theory at a generic point in the $(T,\mu)$ plane is strongly coupled, one has to resort to lattice simulations. It is, however, difficult to carry out Monte Carlo simulations in a theory with a chemical potential due to the so called ``sign problem''. When a theory admits a nonzero chemical potential, the fermionic determinant becomes generally complex, and large oscillations of the phase make the Monte Carlo simulation practically impossible \cite{deForcrand:2010ys}. 

We could expect that the sign problem simplifies in the 't Hooft large-$N$ limit of QCD  \cite{McLerran:2007qj}. In this limit, fundamental matter is quenched, and the hope is that the problem which arises from the determinant is a subleading $1/N$ effect. The problem was first discussed at the diagrammatic level by Cohen in Ref. \cite{Cohen:2004mw}. A related idea which was proposed recently is the large-$N$ equivalence between QCD and other theories which admit a real fermionic determinant \cite{Cherman:2010jj,Hanada:2011ju,Cherman:2011mh,Hanada:2012nj}.
 
In this paper we use the worldline formalism \cite{Strassler:1992zr} to study the dependence of observables on the chemical potential in large-$N$ QCD. In this approach, the fermionic determinant is written as a sum of Wilson (or Polyakov) loops. Moreover, the expansion in loops corresponds to a systematic expansion of the fermionic determinant in powers of $1/N$ \cite{Armoni:2008jy}. The main disadvantage of this approach is that it is perturbative in $1/N$, and hence the expansion is around the Yang-Mills vacuum (without quarks). As a result, the discussion will be restricted to the phases of the theory with small $\mu$ where there is no breaking of baryon (or isospin) number.

The main goal of the paper is to understand how observables depend
on the chemical potential in the large-$N$ theory. We will also
discuss the ``sign problem''. We will explicitly decompose the determinant into real and imaginary parts and shed
light on the origin of the problem. The outcome of our analysis is
that while the large-$N$ theory admits a sign problem, the problem is
not as severe as in the finite-$N$ case. More precisely, we will show
that to leading fermionic contribution in $N$, C-even (real) operators do not couple to the
phase of the fermionic determinant. In other words, for
C-even operators, the sign problem is a subleading effect.

This understanding will be then used to clarify the meaning of the
recently proposed large-$N$ equivalence between QCD theory with a
baryon chemical potential and a QCD theory with an isospin chemical
potential \cite{Hanada:2011ju,Hanada:2012nj}.

We conclude the paper with a discussion about the implications of our analysis on the finite-$N$ theory.

\section{The $1/N$ expansion via the worldline formalism}

Let us review the worldline formalism  \cite{Strassler:1992zr,Armoni:2008jy}. Consider $SU(N)$ gauge theory with one flavor and without a chemical potential. 
We formulate the theory on $R^3 \times S^1$, where $S^1$ is a temporal circle with a radius $R=1/2\pi T$. The basic idea is to express the fermionic determinant in terms of Wilson loops. The precise relation between the {\it Euclidean} fermionic determinant and Wilson loops is as follows:
\begin{equation}
 \det \left ( \slsh \!D +m \right )  = \exp \Gamma [A] \, ,
\end{equation}
and 
\begin{equation}
{\cal Z} = \int {\cal D}A_{\mu} \exp (-S_{\rm YM}) \exp (\Gamma [A]) \, \label{partition} ,
\end{equation}
where
\begin{eqnarray}
\label{wlineint}
 \Gamma [A] &=&
-{1\over 2} \int _0 ^\infty {dt \over t} \exp (-m^2 t)
\nonumber\\[3mm]
 &\times&
\int {\cal D} x {\cal D}\psi
\ (-1)^\omega
\, \exp
\left\{ -\int _0 ^t d\tau \, \left ( {1\over 2} \dot x ^\mu \dot x ^\mu + {1\over
2} \psi ^\mu \dot \psi ^\mu \right )\right\}
\nonumber \\[3mm]
 &\times &  {\rm Tr }\,
{\cal P}\exp \left\{   i\int _0 ^t d\tau
\,  \left (A^a_\mu \dot x^\mu -\frac{1}{2} \psi ^\mu F^a_{\mu \nu}  \psi ^\nu
\right )t^a \right\}  \, ,
\end{eqnarray}
with $x^\mu (0)=x^\mu (t)$ and $\omega$ being the winding number of the path $x_\mu$ around the compact direction. $\{ t^a \} $ is a set of matrices which transform in the fundamental representation of the $SU(N)$ Lie algebra. The factor $(-1)^\omega$ implements the antiperiodic boundary conditions for fermions in the worldline formalism. Thus $\Gamma $, is a sum over (super-)Wilson loops. The sum is over contours of all sizes and shapes.

Let us introduce a chemical potential $\mu$:
\begin{equation}
 \det \left ( \slsh \!D +m \right ) \rightarrow  \det \left ( \slsh \!D  + m+ \mu \gamma ^0 \right ) \, .
\end{equation}
 In the worldline formalism, a chemical potential is easily implemented by introducing a constant background $U(1)$ gauge connection
\begin{equation}
\langle A_\nu \rangle = -i\delta ^0 _\nu  \mu \, ,
\end{equation}
resulting in the introduction of a factor
\begin{equation}
\exp (\mu \int dx^0) = \exp ({\mu \over T} \omega)
\end{equation}
since the integral along the $x^0$ direction (the circle) measures the winding $\omega$. Let us decompose $\Gamma$ accordingly to the various topological sectors which correspond to the winding number
\begin{equation}
 \Gamma {(\mu)} = \sum _{\omega =-\infty} ^{\infty} \Gamma ^{(\omega)} {(\mu)} = 
  \sum _{\omega =-\infty} ^{\infty} \Gamma ^{(\omega)}  {(\mu=0)} \exp ({\mu \over T} \omega) \,.
 \label{gammamu}
\end{equation}
While we cannot prove that the above expansion \eqref{gammamu} is always convergent, it is expected to converge when $m/\mu$ is large enough, since the quark mass serves as an
IR regulator such that each closed loop carries a factor $\sim \exp
(-{m\over T} |\omega|)$.

The free energy at generic $N$ is defined as
\begin{equation}
F = -T \log \mathcal{Z} = F_{\rm YM} - T \log \langle e^{\Gamma} \rangle_{\rm YM} \,.
\end{equation}
We wish to consider now the expansion of $\exp \Gamma$. As was noted in Ref. \cite{Armoni:2009jn}, the connected $k$-point function of $\Gamma$ is suppressed by powers of $N$:
\begin{equation}
\langle \underbrace{\Gamma \Gamma ... \Gamma }_{\rm k\, times} \rangle _c \sim {1\over N ^{(k-2)}}
\end{equation}
Therefore, the expansion of  $\exp \Gamma$ in powers of $\Gamma$ is an expansion in $1/N$. The approximation $\exp \Gamma \sim 1$ corresponds to quenching. In order to consider the effect of fermionic matter, one has to go beyond the quenched approximation, namely, to consider one Wilson loop (which corresponds to one power of $\Gamma$)
\begin{equation}
\exp \Gamma \sim 1 + \Gamma \,.
\end{equation}
In particular, the leading ${\cal O}(N)$ contribution of fermions to the free energy 
is
\begin{equation}
F_f =  - T \log \langle e^{\Gamma} \rangle_{\rm YM} \sim
- T \langle \Gamma \rangle _{\rm YM}=
- T \sum _{\omega =-\infty} ^{\infty}  \exp ({\mu \over T}\omega) \langle \Gamma ^{(\omega)}  {(\mu=0)}  \rangle _{\rm YM} \,.
\label{free}
\end{equation}
Note that the expectation value is calculated in the pure Yang-Mills vacuum.
Next, we consider a generic Wilson loop $O^{(\omega ')}$ with winding number $\omega '$. At ${\cal O}(N)$, the expectation values of  $O^{(\omega ')}$ are given by
\begin{equation}
\langle O^{(\omega ')} \rangle _{\rm QCD} = \langle O^{(\omega ')} \rangle _{\rm YM} + \sum _{\omega =-\infty} ^{\infty} \exp ({\mu \over T} \omega) \langle  O^{(\omega ')} \Gamma ^{(\omega)}  {(\mu=0)} \rangle _{\rm YM}  \,.
\label{observable} 
\end{equation}
In the next sections, we will elaborate on the implications of the expression \eqref{observable}. 

\subsection{$\mu$ dependence in the confining phase}

Let us restrict our attention to the hadronic confining phase. In the confining phase, the expectation value of any loop wrapping $\omega$ times around the circle vanishes unless $\omega \bmod N =0$. At the leading fermionic contribution in $N$, we can neglect ${\cal O}(N)$ windings. Therefore, the free energy \eqref{free} will get a contribution only from the sector with $\omega=0$,
\begin{equation}
F_f = -T  \langle \Gamma ^{(\omega=0)}  {(\mu=0)}  \rangle_{\rm YM} \,.
\label{free1}
\end{equation}
Thus, we conclude that in the hadronic phase the leading fermionic large-$N$ contribution to the quantity $F/T$ is $\mu$ and temperature {\it independent},
\begin{equation}
\frac{1}{T} F_f(T,\mu) = \left. \frac{1}{T} F_f(T,\mu) \right|_{T=0,\mu=0} \,.
\end{equation}
The same conclusion holds for any Wilson loop with a zero winding number,
\begin{equation}
\langle O^{(\omega '=0)}\rangle(T,\mu) = \langle O^{(\omega '=0)} \rangle (T=0,\mu =0) \, .
\end{equation} 

Let us consider a Wilson loop $O^{(\omega ')}$ with $\omega ' \neq 0$. In this case Eq. \eqref{observable} is saturated by $\omega ' + \omega =0$, namely,
\begin{equation}
\langle O^{(\omega ')} \rangle _{\rm QCD}= \langle O^{(\omega ')} \rangle _{\rm YM} + \exp (-{\mu \over T}\omega ') \langle  O^{(\omega ')} \Gamma ^{(-\omega ')}  {(\mu=0)} \rangle  _{\rm YM} \,.
\end{equation}
Even in the confining phase, expectation values of Wilson loops with a nonzero winding admit a nontrivial dependence on the chemical potential.

\subsection{$\mu$ dependence in the deconfining phase}

In the deconfining phase, the Polyakov loop expectation value is nonvanishing. Therefore the free energy is not saturated by a Wilson loop with $\omega =0$, but it rather acquires contributions from all the topological sectors. In particular, the free energy is $\mu$ and temperature {\it dependent}. 
 Using Eq. \eqref{free} and charge-conjugation invariance of the Yang-Mills vacuum, we arrive at
\begin{equation}
F_f = - T \langle \Gamma ^{(0)} (\mu=0) \rangle_{\rm YM} - 2 T \sum _{\omega >0 }  \langle \Gamma ^{(\omega)} (\mu=0) \rangle_{\rm YM}  \cosh ({\mu \over T}\omega) \,.
\label{free2} 
\end{equation}
In general, observables in the deconfining phase are both $\mu$ and temperature dependent.

\subsection{A comment on mesonic correlators}

An important question is how the mesonic spectrum is affected by the chemical potential. For simplicity, we will consider the case of a theory with a single flavor. Our discussion applies also to a multiflavor theory with a baryon chemical potential. 

The mesonic spectrum of a large-$N$ theory is extracted from a two-point function as follows
\begin{equation}
\langle \bar \Psi \Psi (x) \, , \,  \bar \Psi \Psi (y) \rangle = \int {d^4 k \over (2\pi)^4} \exp (i k(x-y)) G(k^2) \label{twopoint}
\end{equation}  
with
\begin{equation}
G(k^2) =\sum _n {f_n^2 \over k^2+ M^2_n} \, ,
\end{equation}
where $f_n$ are coupling constants and $M_n$ are the meson masses. In order to compute the two-point function \eqref{twopoint} in the worldline formalism, one has to add to the action a source $\int d^4 x \, J(x)  \bar \Psi \Psi (x)$ and to differentiate the partition function twice with respect to the current $J$. The outcome \cite{Armoni:2011qv}, to leading order in $N$, is that Eq. \eqref{twopoint} is given by a summation over all Wilson loops which pass via $x$ and $y$,
\begin{equation}
\langle \bar \Psi \Psi (x) \, , \,  \bar \Psi \Psi (y) \rangle = \langle \Gamma _{x,y} (\mu) \rangle \,.
\end{equation} 
Each Wilson loop in the sum carries a factor of $\exp(\mu \omega/T)$ where $\omega$ is its winding number.

However, in the confining phase, the correlator is saturated by Wilson loops with a zero winding number: hence, we conclude that the leading large-$N$ meson spectrum is not affected by the presence of the chemical potential.

\section{The origin of the sign problem}

In this section, we elaborate on the origin of the sign problem in the worldline
 approach.

Let us start by proving the following lemma, valid at fixed gauge configuration:
\begin{equation}
\Gamma ^{(\omega)} (\mu=0) ^\star = \Gamma ^{(-\omega)} (\mu=0)
\label{conjugate}
\end{equation}

Proof: consider the theory with a pure imaginary chemical potential. In this case,
\begin{equation}
\sum _{\omega = -\infty} ^{\infty} \Gamma ^{(\omega)}(i\mu) =
\sum _{\omega = -\infty} ^{\infty} \Gamma ^{(\omega)}(0) \exp (i {\mu \over T} \omega)=
\log  \det \left ( \slsh \!D +m+ i\mu \gamma^0 \right ) \,,
\end{equation}
hence,
\begin{equation}
\Gamma ^{(\omega)}(0) = \int_0^{2 \pi T} {d\mu \over 2 \pi T} \exp (-i {\mu \over T} \omega) \log  \det \left ( \slsh \!D +m+ i\mu \gamma^0 \right )  \, . \label{complex}
\end{equation}
Since the determinant is positive (hence, the logarithm is real), taking the complex conjugate is equivalent to replacing $\omega \to -\omega$. Equation \eqref{complex} can be written also in the following form:
\begin{eqnarray}
& & \Gamma ^{(\omega)}(0)+ \Gamma ^{(-\omega)}(0)  = 2\int {d\mu \over 2\pi T}  \cos ( {\mu \over T} \omega) \log  \det \left ( \slsh \!D+m + i\mu \gamma^0 \right ) \\
& & \Gamma ^{(\omega)}(0) - \Gamma ^{(-\omega)}(0)  = -2i \int {d\mu \over 2\pi T} \sin ( {\mu \over T} \omega) \log  \det \left ( \slsh \!D +m+ i\mu \gamma^0 \right ) \,,
\end{eqnarray}
which separates the real and imaginary parts. In particular, note that $\Gamma ^{(0)}(0)$ is real.

The above analysis has implications on the origin of the sign problem. When $\mu=0$, the fermionic determinant is real because we sum over all configuration with all winding
\begin{equation}
 \det \left ( \slsh \!D  +m \right )  = \exp \sum _{\omega =-\infty} ^{\infty} \Gamma ^{(\omega)}(\mu=0) \, ,
\end{equation}
and in spite of $\Gamma ^{(\omega)}(\mu=0)$ being complex, the imaginary part cancels between  $\Gamma ^{(\omega)}$ and   $\Gamma ^{(-\omega)}$. However, when we introduce a real chemical potential, $\Gamma ^{(\omega)}$ is weighted by $\exp ({\mu \over T} \omega)$, while  $\Gamma ^{(-\omega)}$ is weighted by    $\exp (-{\mu \over T} \omega)$, and the imaginary part no longer cancels:
\begin{equation}
\sum _{\omega =-\infty} ^{\infty} \Gamma ^{(\omega)}(\mu) =
\Gamma ^{(0)}(0) + \sum _{\omega >0} \left ( \exp ({\mu \over T} \omega ) \Gamma ^{(\omega)}(0) +  \exp (-{\mu \over T} \omega ) \Gamma ^{(-\omega)}(0) \right )
\end{equation}
This is the reason for the sign problem. From Eq.\eqref{conjugate}, we learn that
\begin{eqnarray}
& & \Re{(\Gamma ^{(\omega)})} = \Re{(\Gamma ^{(-\omega)})} \\
& & \Im{(\Gamma ^{(\omega)})} = -\Im{(\Gamma ^{(-\omega)})} \, , 
\end{eqnarray}
and the fermionic determinant becomes
\begin{eqnarray}
\det \left ( \slsh \!D  +m +\mu \gamma ^0 \right )  &=& 
\exp \left\{
\Gamma ^{(0)}(0) + 2 \sum _{\omega >0} \cosh ({\mu \over T} \omega ) \Re{(\Gamma ^{(\omega)}(0))} \right\}
\times \nonumber \\
&& \times
\exp \left\{
2i \sum _{\omega >0} \sinh ({\mu \over T} \omega ) \Im{(\Gamma ^{(\omega)}(0))}  \right\} \,. \label{detomega}
\end{eqnarray}
In particular, note that the notorious phase which causes the sign problem is given by
\begin{equation}
 2 \sum _{\omega >0} \sinh ({\mu \over T} \omega ) \Im{(\Gamma ^{(\omega)}(0))} \,.
\end{equation}
In the large-$N$ limit, in expectation values of observables with a well-defined large-$N$ limit, the fermionic determinant can be replaced by
\begin{eqnarray}
& &\det \left ( \slsh \!D  +m +\mu \gamma ^0 \right ) \rightarrow 1 + \Gamma ^{(0)}(0)+ \nonumber \\
& & + \sum _{\omega >0} \left ( 2 \cosh ({\mu \over T} \omega ) \Re{(\Gamma ^{(\omega)}(0))}+ 2i \sinh ({\mu \over T} \omega ) \Im{(\Gamma ^{(\omega)}(0))}  \right ) \,.
\label{bar_det_largeN}
\end{eqnarray}
We want to investigate now for which observables the imaginary part in the previous equation does not contribute and can be dropped.

The free energy in the confining (hadronic) phase is saturated by $\Gamma ^{(0)}(\mu)$, which is real, so obviously there is no sign problem in this case: only configurations with zero winding number contribute to the free energy. A more interesting case is the deconfining phase. The imaginary part in Eq. \eqref{bar_det_largeN} drops once we insert it into Eq.~\eqref{free2}, because the Yang-Mills vacuum is invariant under charge-conjugation ($\langle \Gamma ^{(\omega)}(0) \rangle_{\rm YM} $ is real).

Let us consider now observables which are Wilson loops of arbitrary shape with generic winding number $\omega'$. At the first leading fermionic contribution:
\begin{eqnarray}
\langle O^{(\omega ')} \rangle
&=& \langle O^{(\omega ')} \rangle_{\rm YM} + 2 \sum_{\omega =-\infty} ^{\infty} \exp ({\mu \over T} \omega) \langle  O^{(\omega ')} \Gamma ^{(\omega)}  {(\mu=0)} \rangle_{\rm c,YM} = \nonumber \\
&=& \langle O^{(\omega ')} \rangle_{\rm YM} + 2 \sum_{\omega =-\infty} ^{\infty} \textrm{cosh} ({\mu \over T} \omega) \langle  \Re O^{(\omega ')} \ \Re \Gamma ^{(\omega)}  {(\mu=0)} \rangle_{\rm c,YM} + \nonumber \\
&& - 2 \sum_{\omega =-\infty} ^{\infty} \textrm{sinh} ({\mu \over T} \omega) \langle  \Im O^{(\omega ')} \ \Im \Gamma ^{(\omega)}  {(\mu=0)} \rangle_{\rm c,YM}
\,.
\end{eqnarray}
In the confining phase and for Wilson loops with no winding $\omega'=0$, only $\Gamma ^{(0)}(\mu)$, which is real, contributes because of center symmetry. In the deconfined phase, the imaginary part of the determinant generally contributes.

However it is clear that the imaginary part of the determinant couples only to the imaginary (or C-odd) part of $O^{(\omega ')}$. If one consider only the C-even part, then the imaginary part of the determinant decouples from any generic connected expectation value (in both phases), thanks to the C-invariance of the YM vacuum:
\begin{eqnarray}
& & \langle \Re  O_1^{(\omega_1')} \cdots \Re  O_n^{(\omega_n')} \rangle _{\rm c} = \langle \Re  O_1^{(\omega_1')} \cdots \Re  O_n^{(\omega_n')} \rangle_{\rm c,YM} + \nonumber \\
& & 2 \sum_{\omega =-\infty} ^{\infty} \textrm{cosh} ({\mu \over T} \omega) \langle  \Re  O_1^{(\omega_1')} \cdots \Re  O_n^{(\omega_n')} \ \Re \Gamma ^{(\omega)}  {(\mu=0)} \rangle_{\rm c,YM}
\,.
\end{eqnarray}
The phase of the determinant decouples from $\Re  O^{(\omega ')}$ at
the leading fermionic contribution, which is the first subleading
contribution in $1/N$. The sign problem is therefore a $1/N^2$ effect for $\Re  O^{(\omega ')}$.

In the confining phase only, the imaginary part of the determinant decouples also from the loop $O^{(0)}$ with zero winding number, since only the sector with zero winding contributes (because of center symmetry), and this is $\mu$-independent:
\begin{equation}
\langle O_1^{(0)} \cdots O_n^{(0)} \rangle _{\rm c}  \langle O_1^{(0)} \cdots O_n^{(0)} \rangle_{\rm c,YM} + 2 \langle  O_1^{(0)} \cdots O_n^{(0)} \Gamma ^{(0)}  {(\mu=0)} \rangle_{\rm c,YM}
\,.
\end{equation}

Summarizing, the sign problem does not exist at the leading fermionic contribution in $1/N$ for a class of observables which include real parts of Wilson loops with any winding in both phases, and Wilson loops with no winding in the confining phase only.

\section{Large-$N$ equivalence between baryon and isospin chemical potentials}

Let us consider an $SU(N)$ gauge theory with two identical massive flavors. Let us assign the same chemical potential $\mu$ to both flavors (``baryon chemical potential''). After an integration of the Fermi fields, we obtain a square of the single-flavor fermionic determinant,
\begin{equation}
 \left ( \det  ( \slsh \!D  +m +\mu \gamma ^0 ) \right )^2 \,.
\end{equation}
In the worldline formalism, implementing two identical flavors is achieved, with respect to the single flavor, by the assignment
\begin{equation}
\exp (\Gamma (\mu)) \rightarrow \exp (2\Gamma (\mu)) \,.
\end{equation}
Alternatively, one can assign a chemical potential $\mu$ to one flavor and a chemical potential $-\mu$ to the other flavor (``isospin chemical potential''). In this case, we obtain a product of fermionic determinants,
\begin{equation}
 \left ( \det  ( \slsh \!D  +m +\mu \gamma ^0 ) \right ) \left ( \det  ( \slsh \!D  +m -\mu \gamma ^0 ) \right ) \,,
\end{equation}
resulting in
\begin{equation}
\exp (\Gamma (\mu)) \rightarrow \exp (\Gamma (\mu) + \Gamma (-\mu)) \,.
\end{equation}
It has recently been argued \cite{Hanada:2011ju,Hanada:2012nj} that the above two theories are planar-equivalent in the hadronic and deconfining phases. This means that a \textit{common sector} of observables exits in the two theories, such that all the connected expectation values of such observables are identical in the large-$N$ limit in the two theories. An earlier perturbative argument in favor of the equivalence was given in Ref. \cite{Cohen:2004mw}. The equivalence and its breaking at smaller $N$ are also supported by a recent numerical lattice simulation \cite{Cea:2012ev}.  

Considering an isospin chemical potential instead of a baryon chemical
potential amounts to dropping the imaginary part of the determinant
\eqref{bar_det_largeN} at the leading nontrivial order in
$1/N$. Indeed, the fermionic determinant of a two-flavor theory with a
baryon chemical potential is given by
\begin{eqnarray}
& & \left ( \det  ( \slsh \!D  +m +\mu \gamma ^0  ) \right )^2 \rightarrow 1 + 2\Gamma ^{(0)}(0)+ \nonumber \\
& & + 2\sum _{\omega >0} \left ( 2 \cosh ({\mu \over T} \omega )
  \Re{(\Gamma ^{(\omega)}(0))}+ 2i \sinh ({\mu \over T} \omega )
  \Im{(\Gamma ^{(\omega)}(0))}  \right ) \, ,
\end{eqnarray}
while the fermionic determinant of a two-flavor theory with an isospin
chemical potential is given by
\begin{eqnarray}
& &\det  ( \slsh \!D  +m +\mu \gamma ^0 ) \det  ( \slsh \!D  +m -\mu
\gamma ^0 ) \rightarrow 1 + 2\Gamma ^{(0)}(0)+ \nonumber \\
& & + 2\sum _{\omega >0} \left ( 2 \cosh ({\mu \over T} \omega ) \Re{(\Gamma ^{(\omega)}(0))} \right ) \,.
\end{eqnarray}

Therefore, all the observables analyzed in the previous section which
decouple from the imaginary part of the determinant with baryonic
chemical potential belong to the common sector. 
Put differently, the
common sector of the two theories includes gluonic C-even operators. For
these operators, the sign problem is a subleading effect -- they do not
couple to the phase of the fermionic determinant (which
exists only for the theory with a baryon chemical potential) at the
leading fermionic contribution. In addition to these C-even operators,
there
exist other quantities which belong to the common sector: for example,
the charge density which is obtained by differentiating the free
energy with respect to the chemical potential. Obviously, derivatives
with respect to $\mu$ of quantities which belong to the common sector
also belong to the common sector.

The outcome of the above analysis is that in the large-$N$ limit of
QCD, in a limited part of the phase diagram and for certain operators, it is justified to
replace the theory with a baryon chemical potential by a theory with an isospin chemical potential. 

\section{Discussion}

In this paper we focused on the large-$N$ limit of a QCD theory with a chemical potential. 

Let us discuss what happens when we introduce higher $1/N$ corrections.
The simplest case is the free energy in the hadronic phase,
\begin{equation}
F/T = F_{\rm YM}/T-\log \left ( \sum _{k=0} ^{\infty} {1\over k!} \sum _{l=-\infty} ^{\infty} \exp ({\mu \over T} lN) \langle \Gamma ^{(\omega_1)} \Gamma ^{(\omega_2)}... \Gamma ^{(\omega_k)} \rangle \right )
\end{equation}
with $\omega _1 + \omega _2 + ... + \omega _k = lN$, where $l$ is an
integer. Namely, for a given $k$, one has to consider all the
partitions of $N$, $2N$, $3N$ and so on. These configurations
contribute to the free energy and lead to a nontrivial $\mu$
dependence\cite{Bringoltz:2010iy,Gattringer:2009wi}. For this reason,
our conclusions about the large-$N$ theory cannot apply for the
finite-$N$ theory. 

Another important issue is the sign problem at finite $N$. At
infinite $N$, for C-even observables, only the real part of $\Gamma$
contributes: hence, $\exp \Gamma$ is positive, and the measure is positive-definite. This is not the case at
finite $N$ and the sign problem is severe.

The main result of our paper is a field-theory understanding of the planar equivalence between the theory with a baryon chemical potential and the theory with an isospin chemical potential. It is now clear {\it why} it is useful to carry out lattice simulations using an isospin chemical potential: there is a limit of QCD (the large-$N$ limit) where this procedure is justified. At finite $N$, and, in particular, for $SU(3)$, we expect an error of $1/N$ in general, and an error of $1/N^2$ for purely gluonic C-even observables.

Among future directions of investigation we would like to mention planar
equivalence with $SO(N)$ theories and theories with adjoint matter. It
is straightforward to generalize our discussion to these cases. 

\vskip 1cm

{\it \bf Acknowledgements.} We thank G. Aarts, A. Cherman, S. Hands, C. Hoyos, P. Kumar, B. Lucini, H. Neuberger and L. Yaffe for discussions. A.A. would like to thank CERN for the hospitality where this work began.



\begin{thebibliography}{99}

\bibitem{Stephanov:2007fk} 
  M.~A.~Stephanov,
  ``QCD phase diagram: An Overview,''
  PoS LAT {\bf 2006}, 024 (2006)
  [hep-lat/0701002].

\bibitem{deForcrand:2010ys} 
  P.~de Forcrand,
  ``Simulating QCD at finite density,''
  PoS LAT {\bf 2009}, 010 (2009)
  [arXiv:1005.0539 [hep-lat]].

\bibitem{McLerran:2007qj} 
  L.~McLerran and R.~D.~Pisarski,
  ``Phases of cold, dense quarks at large N(c),''
  Nucl.\ Phys.\ A {\bf 796}, 83 (2007)
  [arXiv:0706.2191 [hep-ph]].

\bibitem{Cohen:2004mw} 
  T.~D.~Cohen,
  ``Large N(c) QCD at non-zero chemical potential,''
  Phys.\ Rev.\ D {\bf 70}, 116009 (2004)
  [hep-ph/0410156].

\bibitem{Cherman:2010jj} 
  A.~Cherman, M.~Hanada and D.~Robles-Llana,
  ``Orbifold equivalence and the sign problem at finite baryon density,''
  Phys.\ Rev.\ Lett.\  {\bf 106}, 091603 (2011)
  [arXiv:1009.1623 [hep-th]].

\bibitem{Hanada:2011ju} 
  M.~Hanada and N.~Yamamoto,
  ``Universality of Phases in QCD and QCD-like Theories,''
  JHEP {\bf 1202}, 138 (2012)
  [arXiv:1103.5480 [hep-ph]].

\bibitem{Cherman:2011mh} 
  A.~Cherman and B.~C.~Tiburzi,
  ``Orbifold equivalence for finite density QCD and effective field theory,''
  JHEP {\bf 1106}, 034 (2011)
  [arXiv:1103.1639 [hep-th]].

\bibitem{Hanada:2012nj} 
  M.~Hanada, C.~Hoyos, A.~Karch and L.~G.~Yaffe,
  ``Holographic realization of large-Nc orbifold equivalence with non-zero chemical potential,''
  arXiv:1201.3718 [hep-th].

\bibitem{Strassler:1992zr}
  M.~J.~Strassler,
  ``Field theory without Feynman diagrams: One loop effective actions,''
  Nucl.\ Phys.\  B {\bf 385}, 145 (1992)
  [arXiv:hep-ph/9205205].

\bibitem{Armoni:2008jy}
  A.~Armoni,
  ``Beyond The Quenched (or Probe Brane) Approximation in Lattice (or Holographic) QCD,''
  Phys.\ Rev.\  D {\bf 78}, 065017 (2008)
  [arXiv:0805.1339 [hep-th]].

\bibitem{Armoni:2009jn} 
  A.~Armoni,
  ``The Conformal Window from the Worldline Formalism,''
  Nucl.\ Phys.\ B {\bf 826}, 328 (2010)
  [arXiv:0907.4091 [hep-ph]].

\bibitem{Armoni:2011qv} 
  A.~Armoni and O.~Mintakevich,
  ``Comments on Mesonic Correlators in the Worldline Formalism,''
  Nucl.\ Phys.\ B {\bf 852}, 61 (2011)
  [arXiv:1102.5318 [hep-th]].

\bibitem{Cea:2012ev} 
  P.~Cea, L.~Cosmai, M.~D'Elia, A.~Papa and F.~Sanfilippo,
  ``The critical line of two-flavor QCD at finite isospin or baryon densities from imaginary chemical potentials,''
  arXiv:1202.5700 [hep-lat].


\bibitem{Bringoltz:2010iy} 
  B.~Bringoltz,
  ``Large-N spacetime reduction and the sign and silver-blaze problems of dense QCD,''
  JHEP {\bf 1006}, 076 (2010)
  [arXiv:1004.0030 [hep-lat]].

\bibitem{Gattringer:2009wi} 
  C.~Gattringer and L.~Liptak,
  ``Canonical fermion determinants in lattice QCD: Numerical evaluation and properties,''
  Phys.\ Lett.\ B {\bf 697}, 85 (2011)
  [arXiv:0906.1088 [hep-lat]].


\end{thebibliography}
\end{document}